\documentclass[conference]{IEEEtran}
\IEEEoverridecommandlockouts
\usepackage{cite}
\usepackage{amsmath,amssymb,amsfonts}
\usepackage{algorithmic}
\usepackage{graphicx}
\usepackage{textcomp}
\usepackage{xcolor}
\usepackage{url}
\usepackage[hidelinks]{hyperref}
\def\BibTeX{{\rm B\kern-.05em{\sc i\kern-.025em b}\kern-.08em
    T\kern-.1667em\lower.7ex\hbox{E}\kern-.125emX}}
\begin{document}

\title{A Secure, Manifest-Based Framework for Delegated Privilege Promotion}

\author{\IEEEauthorblockN{1\textsuperscript{st} Rajarshi Chowdhury}
\IEEEauthorblockA{\textit{Oracle America Inc.} \\
Redwood Shores, CA, USA \\
rajarshi.chowdhury@oracle.com}
\and
\IEEEauthorblockN{2\textsuperscript{nd} Akshay Shah}
\IEEEauthorblockA{\textit{Oracle America Inc.} \\
Redwood Shores, CA, USA \\
akshay.shah@oracle.com}
\thanks{Accepted at the 56\textsuperscript{th} Annual IEEE/IFIP International Conference on Dependable Systems and Networks (DSN 2026).}
}

\maketitle

\begin{abstract}
Large-scale enterprise software systems commonly run as unprivileged service accounts to enforce least privilege, yet still depend on a small set of privileged components—such as executables with elevated ownership, permissions, or capabilities—for narrowly scoped operations. This creates a persistent security and operational conflict during maintenance. Automated patching tools running without elevated privileges cannot safely update privileged components without either executing the entire patch with full administrative rights or requiring manual administrator intervention. We present a secure, manifest-based infrastructure for delegated promotion of privileged software components, deployed in production as part of a large-scale enterprise database system serving both cloud and on-premises installations. The design centers on a minimal privileged mediator that validates cryptographically protected metadata and allows an unprivileged process to promote only vendor-approved files. The system explicitly mitigates Time-of-Check-to-Time-of-Use (TOCTOU) attacks using file-descriptor–bound validation and promotion, supports offline key rotation and revocation, and enables zero-downtime self-update via atomic replacement.
\end{abstract}


\section{Introduction}

\begin{figure}[t]
  \centering
  \includegraphics[width=0.95\columnwidth]{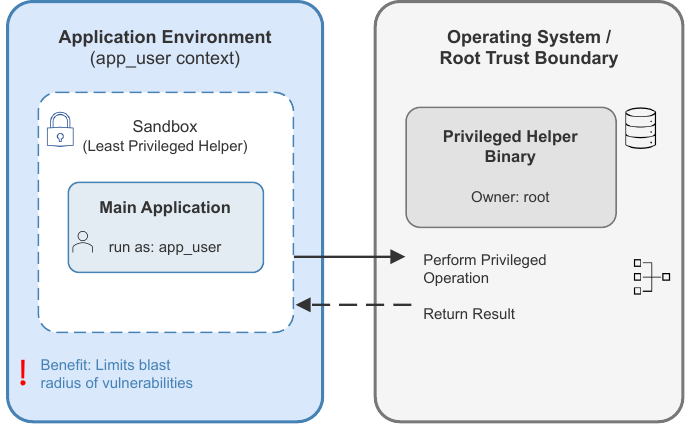}
  \caption{Least-privilege architecture: an unprivileged service account invokes a small set of privileged components for narrowly scoped operations.}
  \label{fig:leastpriv}
\end{figure}

Modern enterprise software systems—including databases, middleware, application servers, and clustered storage platforms—are commonly deployed under unprivileged service accounts to enforce the principle of least privilege and reduce blast radius under compromise. Despite this design, such systems inevitably rely on a small set of privileged components to perform narrowly scoped operations, such as interacting with hardware, modifying kernel-visible state, or accessing protected system resources. These components are typically realized as executables or files with elevated ownership, permissions, or capabilities, and are intentionally limited in scope to keep the trusted computing base small and auditable. Figure~\ref{fig:leastpriv} illustrates this common least-privilege architecture.

This separation between unprivileged services and privileged components breaks down during maintenance. Standard patch utilities run without elevated privileges and therefore cannot safely overwrite root-owned or otherwise protected files. Relaxing permissions to permit unprivileged writes either strips required privilege attributes or introduces a window in which a compromised service account can tamper with components before they are re-privileged. As a result, many deployments fall back to one of two unsatisfactory approaches: executing the entire patching workflow with full administrative privileges, or requiring manual administrator intervention for privileged steps such as changing ownership or invoking a package manager. Both approaches undermine least-privilege guarantees, complicate automation, and introduce human error.

These limitations are particularly acute in modern operational environments. Enterprise software is increasingly deployed via automated CI/CD pipelines, unattended provisioning workflows, and appliance-style installations, including environments that are offline or air-gapped for security, regulatory, or operational reasons. In such settings, reliance on interactive administrative access or network-dependent update mechanisms is impractical, yet privileged components must still be updated safely and reliably.

We present a secure, manifest-based infrastructure for delegated promotion of privileged software components. The system has been deployed in production as part of a large-scale enterprise database, where it automates quarterly updates of privileged helpers across cloud and on-premises installations. The design centers on a minimal privileged mediator that validates cryptographically protected metadata supplied by the software vendor and promotes only those files and privilege attributes explicitly authorized. Validation and promotion are bound to stable kernel objects using file descriptors, eliminating Time-of-Check-to-Time-of-Use (TOCTOU) vulnerabilities inherent in path-based approaches~\cite{bishop,toctou_wei}. The system further supports offline key rotation and revocation in accordance with established key-management practices~\cite{nist80057}, and zero-downtime self-update via atomic replacement.

Existing mechanisms such as sudo-based scripts, policy-driven authorization frameworks like polkit~\cite{polkit}, and system-wide package managers are ill-suited to this problem: they either remain vulnerable to race conditions, require a large trusted computing base, or assume a fully privileged execution model inappropriate for delegated, application-scoped patching. Secure software update frameworks such as TUF~\cite{tuf} address integrity and rollback protection for distributed artifacts, but do not solve the host-local privilege-promotion problem under a compromised unprivileged account.

Our contributions are:
\begin{itemize}
  \item \textbf{Delegated privilege promotion:} An infrastructure that allows an unprivileged patching process to trigger narrowly scoped promotion of privileged components based solely on cryptographically authenticated vendor intent, without granting general administrative authority.
  \item \textbf{Minimal trusted core:} A small privileged mediator that forms the sole executable component of the trusted computing base, supported by root-owned trust anchors and untrusted patch artifacts.
  \item \textbf{Secure lifecycle for offline environments:} Offline-capable key rotation and revocation via a Key Revocation List (KRL), and atomic self-update.
\end{itemize}

\section{System Architecture and Design}

The design centers on a single privileged mediator, referred to as the \emph{enabler}, which enforces all security-critical decisions. The enabler is installed once by an administrator and is the only executable component of this infrastructure that runs with elevated privileges. Its sole function is to validate vendor-authorized update metadata and to promote only those privileged components explicitly approved by that metadata.

\begin{figure}[t]
  \centering
  \includegraphics[width=0.95\columnwidth]{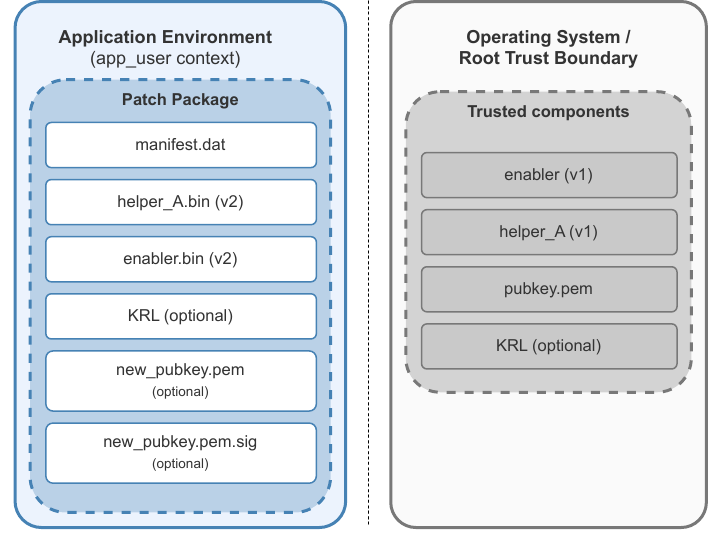}
  \caption{System components: trusted (root-owned) enabler, public key, and key revocation list (KRL); untrusted patch package containing a signed manifest and candidate binaries.}
  \label{fig:components}
\end{figure}

Figure~\ref{fig:components} illustrates the system components and their trust boundaries. The trusted computing base consists exclusively of the root-owned enabler binary and a small set of trust anchors: the vendor public key and a key revocation list (KRL). All patch artifacts, including signed manifests and candidate binaries, are treated as untrusted input and may be fully attacker-controlled.

\subsection{Threat Model}

We assume an attacker has fully compromised the unprivileged service account under which the application and its patching utilities execute. The attacker can read, write, create, delete, and replace files owned by this account, execute arbitrary code as that user, and attempt to exploit race conditions during patch execution. The attacker does not possess root privileges and cannot modify root-owned files, including the enabler binary, the trusted public key, or the KRL.

The attacker's objective is to subvert the patching process to promote a malicious payload as a privileged component (e.g., with elevated ownership, permissions, or capabilities), thereby achieving persistent privilege escalation.

\subsection{Manifest-Based Delegation}

Delegated promotion is driven by a vendor-supplied, cryptographically signed manifest packaged as a single signed envelope within each patch package. The manifest enumerates the privileged actions being authorized and, for each component to be promoted, specifies the candidate file path, destination path, target ownership or permissions (or capabilities), and a cryptographic hash of the file contents.

The envelope is signed using a vendor private key and verified by the enabler using a root-owned public key before any contents are parsed or trusted. The current implementation uses RSA-2048, which is the signing primitive available in our enterprise cryptographic toolchain. It provides 112-bit security, which is adequate for our threat model~\cite{nist80057}. The design is not tied to a specific algorithm: the offline key-rotation mechanism in Section~\ref{sec:keyrotation} is the path to migrate to Ed25519 or a stronger primitive before the NIST SP~800-57 deadline that retires RSA-2048. By treating the signed manifest envelope as the sole authorization source, the system decouples privilege decisions from untrusted patching logic and ensures that promotion is driven exclusively by authenticated vendor intent.

\subsection{Three-Phase Workflow}

The enabler executes a linear, all-or-nothing workflow that separates privileged and unprivileged operations to minimize risk and eliminate race conditions. The workflow consists of three phases:

\begin{enumerate}
  \item Privileged setup:
  The enabler starts with elevated privileges and opens the root-owned trust anchors (public key and KRL), binding validation to stable file descriptors via \texttt{open()} and \texttt{fstat()}. It verifies revocation status and loads trusted metadata into memory, then drops elevated privileges.

  \item Unprivileged validation:
  Running without elevated privileges, the enabler verifies the manifest signature using the in-memory public key and processes each listed component. For each candidate file, it opens the file, computes its cryptographic hash directly from the resulting file descriptor, and retains only descriptors corresponding to successfully verified components.

  \item Privileged promotion:
  After all validation succeeds, the enabler regains elevated privileges. If the manifest authorizes an updated enabler binary, it installs the update and performs a controlled \texttt{execve()} before continuing. The enabler then promotes each verified component by copying bytes from the previously validated file descriptors to their destination paths and applying the authorized ownership, permissions, or capabilities using file-descriptor–safe operations (e.g., \texttt{fchown()}, \texttt{fchmod()}). Any failure aborts execution without partially modifying system state.
\end{enumerate}

\begin{figure}[t]
  \centering
  \includegraphics[width=0.95\columnwidth]{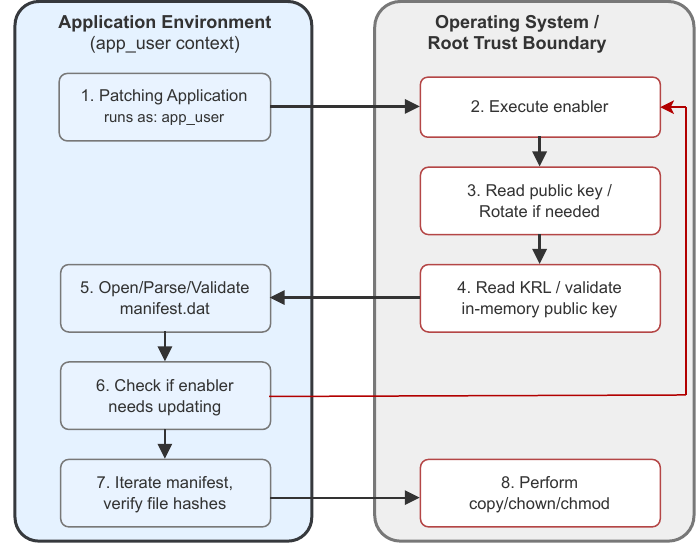}
  \caption{Core validation and promotion workflow: privileged setup, unprivileged validation, and privileged promotion, with security-critical operations bound to verified file descriptors.}
  \label{fig:workflow}
\end{figure}

By separating validation from promotion and binding all security-critical checks and actions to persistent file descriptors rather than filesystem paths (Figure~\ref{fig:workflow}), the enabler guarantees that the content promoted with elevated privileges is exactly the content that was previously validated. This design ensures that no privileged operation is performed on filesystem objects that were not previously validated under the same kernel object identity.

\section{Long-Term Trust Management and Lifecycle}

A deployable privileged-promotion infrastructure must remain secure over time, including under key compromise, software evolution, and offline or air-gapped operation. In addition to validating individual patch packages, the enabler provides explicit mechanisms for cryptographic agility, controlled self-update, and key revocation.

\subsection{Offline Key Rotation}
\label{sec:keyrotation}

The system supports offline rotation of vendor signing keys without requiring network access or interactive administrative intervention. A patch package may include a candidate public key accompanied by a cryptographic authorization generated using the currently trusted private key. During validation, the enabler verifies this authorization using the in-memory trusted public key that was loaded prior to privilege drop.

\begin{figure}[t]
  \centering
  \includegraphics[width=0.75\columnwidth]{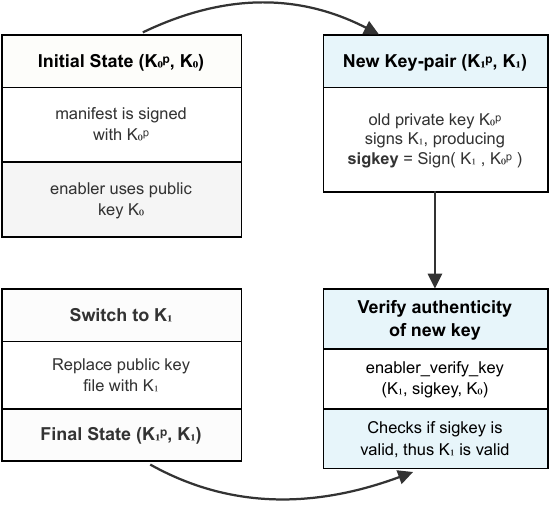}
  \caption{Offline key rotation: a new public key is authorized by a signature from the currently trusted key and atomically installed after successful package authentication.}
  \label{fig:key_rotation}
\end{figure}

Only after the patch package itself has been authenticated under the current trust state does the enabler apply a key update. Installation of a new public key is performed atomically, replacing the existing root-owned key file (e.g., \texttt{pubkey.pem}) using filesystem operations that prevent partial or inconsistent updates. This ensures that key rotation advances trust state monotonically and cannot retroactively legitimize previously unauthenticated packages. Figure~\ref{fig:key_rotation} illustrates this flow.

\subsection{Secure Self-Update and Zero-Downtime Operation}

The enabler itself is treated as a privileged component subject to the same manifest-based authorization as other protected files. If a patch package explicitly authorizes an updated enabler binary, the current enabler validates the new binary, installs it atomically, and performs a controlled \texttt{execve()} into the updated executable before promoting any additional privileged components. This guarantees that all subsequent validation and promotion logic executes under the most recent trusted implementation.

For other privileged helpers, atomic filesystem replacement (e.g., via \texttt{rename()}) enables zero-downtime updates: processes that have already opened or executed an older version continue unaffected, while future invocations transparently use the updated version. This property is critical for maintaining availability during patching of long-running services.

\subsection{Key Revocation}

Signature verification alone is insufficient when a signing key is known or suspected to be compromised. To address this, the system maintains a root-owned Key Revocation List (KRL) that records fingerprints of revoked signing keys.

The KRL is consulted as part of the initial trust establishment phase, prior to authenticating any patch package. A patch is rejected if its signing key is already present in the currently installed KRL. Patch packages may include an updated KRL, but such updates are applied only after the package has been authenticated under the existing trust state. This ordering ensures that revocation state evolves monotonically and that a compromised key cannot legitimize malicious updates by self-revocation.

The KRL revokes keys, not component versions. On its own, it does not stop an attacker from replaying an old but validly-signed manifest whose signing key is still trusted. In production, version-level rollback is an administrative operation handled at the product patchset level: the manifest, trust anchors, and privileged components are rolled back together in one controlled action, not by the enabler on its own. Adding a per-component version counter to the enabler, stored alongside the KRL, is a natural extension and is left as future work.

\section{Security Analysis}

The central security objective of the enabler is to ensure that the bytes promoted with elevated privileges are exactly those that were previously authorized and validated, even when the unprivileged execution context is fully attacker-controlled. The primary threat to this objective arises from Time-of-Check-to-Time-of-Use (TOCTOU) races, in which filesystem objects validated by pathname are replaced before being used in privileged operations.

The enabler eliminates this class of vulnerabilities by binding all security-critical validation and promotion steps to stable kernel objects, specifically file descriptors bound to inodes. We analyze two representative TOCTOU vectors and show how this design prevents privilege subversion in both cases.

\subsection{Race on Trusted Key Material}

A common TOCTOU vulnerability occurs when a privileged program validates a trusted file by pathname and later reopens that path for use. In the enabler's context, such a pattern would allow an attacker controlling the unprivileged filesystem namespace to replace the trusted public key file (e.g., \texttt{pubkey.pem}) between validation and use, thereby subverting signature verification.

\begin{figure}[t]
  \centering
  \includegraphics[width=0.95\columnwidth]{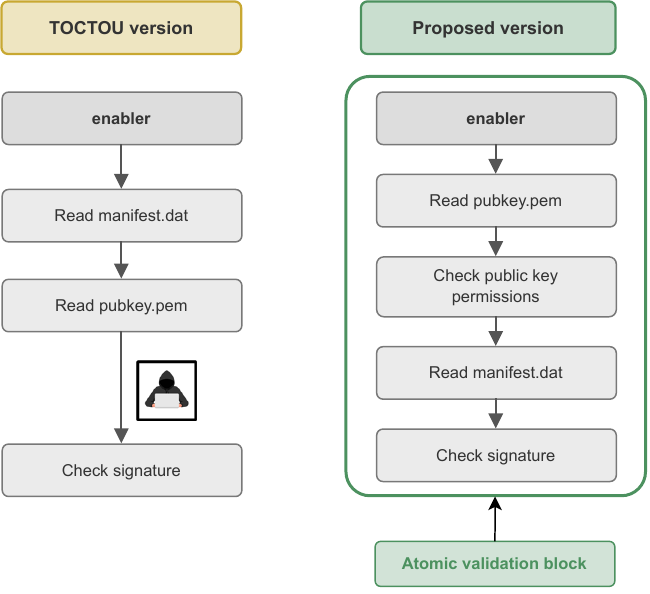}
  \caption{TOCTOU on trusted key material: path-based validation and use is vulnerable to replacement attacks, while FD-bound open, validation, and use is resilient.}
  \label{fig:pubkey_toctou}
\end{figure}

The enabler prevents this attack by opening the trusted public key exactly once during privileged setup and immediately validating its ownership and permissions using \texttt{fstat()} on the resulting file descriptor. The key material is then read exclusively via this descriptor and retained in memory for the remainder of execution. Subsequent renames or replacements at the original path do not affect the already-open descriptor, ensuring that signature verification uses exactly the key that was validated. Figure~\ref{fig:pubkey_toctou} contrasts the vulnerable path-based pattern with the FD-bound approach used by the enabler.

\subsection{Race on Patch Payloads}

An analogous TOCTOU risk exists for unprivileged patch artifacts. If a candidate binary were validated by pathname and later reopened during privileged promotion, an attacker could replace the file between these phases, causing unauthorized content to be promoted.

\begin{figure}[t]
  \centering
  \includegraphics[width=0.95\columnwidth]{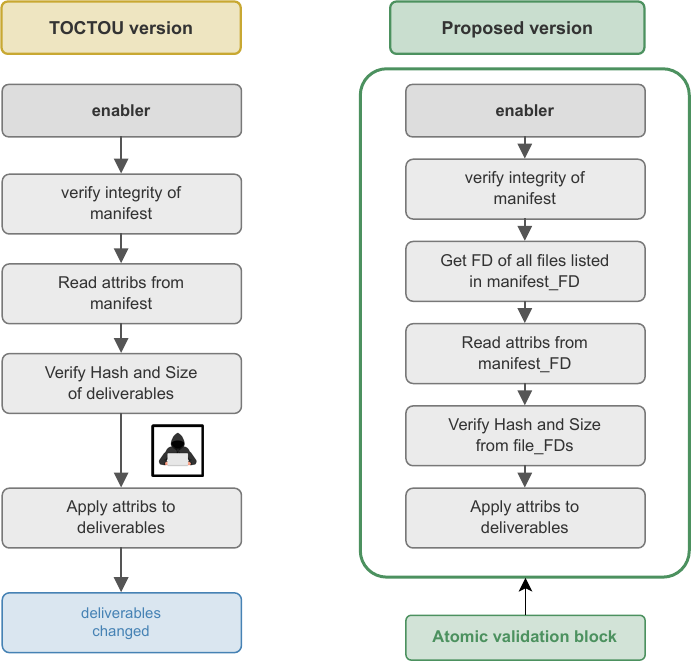}
  \caption{TOCTOU on patch payloads: by binding both validation and promotion to the same file descriptor, path-based replacement attacks do not alter the promoted bytes.}
  \label{fig:payload_toctou}
\end{figure}

The enabler avoids this class of attacks by opening each candidate file during unprivileged validation and computing its cryptographic hash directly from the resulting file descriptor. Only descriptors corresponding to successfully verified files are retained. During privileged promotion, the enabler copies bytes from these same validated descriptors into the destination paths and applies the authorized ownership, permissions, or capabilities. Because both validation and promotion operate on the same descriptor bound to a specific inode, replacing the file at its original path has no effect on the promoted content. Figure~\ref{fig:payload_toctou} shows this difference in flow.

\subsection{Security Invariant}

Together, these mechanisms enforce the following invariant: \emph{no privileged operation is performed on filesystem objects that were not previously validated and authorized under the same kernel object identity}. This invariant holds regardless of attacker-controlled path manipulation and provides the basis for the enabler's resistance to TOCTOU-based privilege escalation.

The invariant follows from the design: (i) every privileged operation in the promotion phase uses a file descriptor as its operand, not a pathname; (ii) each descriptor was opened, and its content cryptographically validated, during the earlier unprivileged validation phase; (iii) in the POSIX model, a file descriptor is bound to the inode that was referenced when \texttt{open()} was called, and does not follow later changes to the pathname; (iv) therefore any rename, unlink, or replacement at the original pathname after validation does not change the descriptor, and does not change the bytes used in the promotion phase. As long as kernel semantics for file descriptors are correct—which the threat model in Section~II-A already assumes—the invariant holds for every component the enabler promotes.

\subsection{Scope and Limitations}

The enabler addresses privilege escalation via TOCTOU races and unauthorized file promotion. It does not protect against vulnerabilities within the privileged helpers themselves, nor does it defend against attacks that compromise root-level access through other vectors. The system assumes the underlying operating system kernel and filesystem behave correctly; kernel-level compromise is outside the threat model.

\section{Implementation and Operational Experience}

\begin{table*}[!t]
\caption{Comparison of Approaches for Updating Privileged Components}
\label{tab:comparison}
\centering
\small
\begin{tabular}{lccccc}
\hline
Approach & Automated & Offline-capable & TOCTOU-safe$^{\mathrm{a}}$ & Minimal TCB & Delegatable$^{\mathrm{b}}$ \\ \hline
sudo rules/scripts & No & Yes & No & No & Partial \\
polkit-based service & Yes & Yes & No & No & Partial \\
System package manager & Yes & Yes & Partial & No & No \\
Enabler (this work) & Yes & Yes & Yes & Yes & Yes \\ \hline
\multicolumn{6}{l}{\footnotesize $^{\mathrm{a}}$ With respect to delegated promotion under an attacker-controlled unprivileged namespace.} \\
\multicolumn{6}{l}{\footnotesize $^{\mathrm{b}}$ Invocable by an unprivileged process to perform narrowly scoped privileged actions without broad administrative authority.} \\
\end{tabular}
\end{table*}

We implemented the enabler as a compact C program using standard POSIX file APIs and OpenSSL for cryptographic operations. The implementation deliberately avoids network access, dynamic policy engines, and complex runtime dependencies. All parsing and cryptographic validation of untrusted inputs is confined to the unprivileged execution phase, while the privileged code path is kept minimal and deterministic.

The trusted computing base consists solely of the enabler binary and a small set of root-owned trust anchors (the vendor public key and Key Revocation List). All patch artifacts, including signed manifest envelopes and candidate binaries, are treated as untrusted input. Privileged operations are restricted to a narrow set of file-descriptor--safe system calls (\texttt{open}, \texttt{fstat}, \texttt{read}, \texttt{rename}, \texttt{fchown}, \texttt{fchmod}), resulting in a short, auditable privileged execution path with a well-defined security boundary.

\subsection{Deployment Experience}

The enabler is deployed in production as part of a large-scale enterprise database system, serving both cloud and on-premises installations. It executes as part of every quarterly update cycle, automating the promotion of privileged helper components that were previously updated through manual administrator intervention or elevated patching scripts. This deployment spans a diverse range of customer environments, including air-gapped and appliance-style configurations where interactive administrative access is impractical.

By integrating the enabler into the standard patch workflow, updates to privileged components now proceed without requiring broad administrative authority for patching tools or manual root intervention, while preserving strict least-privilege boundaries. All changes to the enabler are reviewed internally, and the enabler is exercised by the product's integration test suite before every release.

\subsection{Performance}

On a representative production host, the enabler takes about 62\,ms of wall-clock time to validate and promote a manifest with ten component entries. Of that, roughly 11\,ms is spent in user mode and 31\,ms in kernel mode. Most of the time goes to file I/O and signature verification, and the enabler's overhead is a small part of the overall patch cycle.

\subsection{Availability Considerations}

The enabler runs during patching, not on the request-serving path, so its failure modes affect patch completion rather than steady-state service availability. Even so, they are worth describing. The three-phase workflow (Section~II-C) is designed so that any failure during privileged setup or unprivileged validation aborts the run before any promotion has started, and system state is left unchanged. During the privileged promotion phase, each component is installed by atomic filesystem replacement, so a failure between components never leaves a partially written privileged binary on disk. If the enabler fails for any unexpected reason, administrators can fall back to the traditional root-administered update path to finish the patch manually, and already-running services are not affected.

\subsection{Lessons Learned}

Operational experience highlighted several important lessons. First, path-based validation patterns—common in administrative tooling—are fundamentally incompatible with delegated privilege models: even careful sequencing leaves exploitable race windows under attacker-controlled namespaces. Binding both validation and promotion to stable kernel objects proved essential not only for security, but also for simplifying reasoning about correctness and auditability.

Second, minimizing the privileged code path significantly reduced implementation complexity and review burden. Confining complex parsing and cryptographic operations to the unprivileged phase allowed the privileged mediator to remain small, deterministic, and easier to audit.

Third, designing for graceful fallback was critical for operational confidence. The system preserves the traditional root-administered update path as a fallback mechanism; if the enabler encounters an unexpected failure, administrators retain the ability to apply updates manually. This fallback has never been required in production, but its availability reduced organizational friction during initial adoption.

Finally, the self-update mechanism proved essential for maintaining the enabler itself. Because the enabler validates and installs its own replacement before promoting other components, bugs discovered in the enabler can be corrected in the same patch package that delivers the fix. As long as the defect does not prevent successful validation and \texttt{execve()} into the updated binary, the corrected implementation executes all subsequent promotion steps. This property has allowed the enabler to evolve without requiring out-of-band administrative intervention.

\section{Related Work}

Our work addresses a problem at the intersection of privilege delegation, secure software updates, and TOCTOU mitigation. Common approaches to delegated privilege, such as \texttt{sudo}~\cite{sudo} and policy frameworks like polkit~\cite{polkit}, decouple validation from promotion: files are verified in an unprivileged context and later promoted in a separate privileged step, creating exploitable race windows long recognized as vulnerable to TOCTOU attacks~\cite{bishop,toctou_wei}. Privilege separation techniques~\cite{privsep,openssh_measures} and capability-based confinement~\cite{capsicum,seccomp} reduce the impact of compromise but do not address how an unprivileged process can safely install or re-privilege binaries. Traditional package managers~\cite{rpm,dpkg} execute entirely with administrative privileges and assume full system ownership, making them unsuitable for delegated, application-scoped patching. Atomic update systems like OSTree~\cite{ostree,rpmostree} and Nix~\cite{nix} improve robustness but still assume a privileged update agent. Table~\ref{tab:comparison} summarizes how these approaches compare against the enabler along the axes most relevant to delegated patching of privileged components.

Secure update frameworks such as TUF~\cite{tuf} and Uptane~\cite{uptane_escar,uptane_whitepaper} protect distribution pipelines against key compromise and rollback attacks, while supply-chain systems like in-toto~\cite{intoto}, Sigstore~\cite{sigstore}, and SLSA~\cite{slsa} verify provenance upstream. These ensure correct artifacts reach a host but assume trusted installation contexts and do not address host-local TOCTOU vulnerabilities. Integrity enforcement mechanisms including fs-verity~\cite{fsverity}, dm-verity~\cite{dmverity,avb}, IMA~\cite{ima}, and PRIMA~\cite{prima,ima_subvert} provide strong guarantees for deployed software but are not designed for automated re-privileging of mutable components during patching. Our enabler combines cryptographically authenticated delegation with file-descriptor--bound operations, ensuring that validation and promotion act on identical kernel objects regardless of attacker-controlled path manipulation.

\section{Conclusion}

We presented a secure, manifest-based infrastructure for delegated promotion of privileged software components, resolving a long-standing conflict between least-privilege deployment and automated maintenance. The system combines cryptographically authenticated authorization with file-descriptor--bound TOCTOU mitigation, supports offline key rotation and revocation, and enables controlled self-update—all without requiring broad administrative authority or manual root intervention.

The enabler is deployed in production as part of a large-scale enterprise database system, where it has automated quarterly updates of privileged components across diverse cloud and on-premises environments, reducing operational complexity while strengthening least-privilege boundaries. We believe the approach generalizes to any enterprise software that combines unprivileged service accounts with a small set of privileged helpers—a common pattern across middleware, clustered storage, and appliance-style deployments.

\end{document}